\documentclass[conference]{IEEEtran}
\IEEEoverridecommandlockouts
\usepackage{cite}
\usepackage{subfigure}
\usepackage{amsmath,amssymb,amsfonts}
\usepackage{algorithm,algorithmic}
\usepackage{etoolbox}
\usepackage{graphicx}   
\usepackage{textcomp}
\usepackage{bbold}
\usepackage{xcolor}
\usepackage{amsthm}
\newtheorem{theorem}{Theorem}

\newtheorem{definition}{Definition}
\newtheorem{lemma}{Lemma}
\newtheorem{assumption}{Assumption}

\newcommand{\norm}[1]{\left\lVert#1\right\rVert}
\definecolor{bostonuniversityred}{rgb}{0.8, 0.0, 0.0}

\newcommand{\algorithmicdoinparallel}{\textbf{do} in parallel}
\makeatletter
\AtBeginEnvironment{algorithmic}{%
  \newcommand{\FORP}[2][default]{\ALC@it\algorithmicfor\ #2\ %
    \algorithmicdoinparallel\ALC@com{#1}\begin{ALC@for}}%
}
\makeatother

\def\BibTeX{{\rm B\kern-.05em{\sc i\kern-.025em b}\kern-.08em
    T\kern-.1667em\lower.7ex\hbox{E}\kern-.125emX}}
\begin{document}

\title{Asynchronous Decentralized Learning over Unreliable Wireless Networks}


\author{
    \IEEEauthorblockN{Eunjeong Jeong$^*$, Matteo Zecchin$^*$, and Marios Kountouris}
    \IEEEauthorblockA{Communication Systems Department\\
    EURECOM, Sophia Antipolis, France
    \\\{eunjeong.jeong, matteo.zecchin, marios.kountouris\}@eurecom.fr}
}

\maketitle
\def\thefootnote{*}\footnotetext{These authors contributed equally to this work.\\
The work of E. Jeong is supported from a Huawei France-funded Chair on Future Wireless Networks. The work of M. Zecchin is funded by the Marie Sklodowska Curie action WINDMILL (grant No. 813999).}

\begin{abstract}
Decentralized learning enables edge users to collaboratively train models by exchanging information via device-to-device communication, yet prior works have been limited to wireless networks with fixed topologies and reliable workers. In this work, we propose an asynchronous decentralized stochastic gradient descent (DSGD) algorithm, which is robust to the inherent computation and communication failures occurring at the wireless network edge. We theoretically analyze its performance and establish a non-asymptotic convergence guarantee. Experimental results corroborate our analysis, demonstrating the benefits of asynchronicity and outdated gradient information reuse in decentralized learning over unreliable wireless networks.
\end{abstract}


\section{Introduction}
Distributed learning algorithms empower devices in wireless networks to collaboratively optimize the model parameters by alternating between local optimization and communication phases. Leveraging the aggregated computational power available at the wireless network edge in a communication efficient \cite{mcmahan2017communication} and privacy preserving manner \cite{yan2012distributed}, distributed learning is considered to be a key technology enabler for future intelligent networks.
A promising paradigm, which enables collaborative learning among edge devices communicating in a peer-to-peer (server-less) manner, is decentralized learning \cite{tsitsiklis1986distributed}. Differently from federated learning, decentralized algorithms do not require a star topology with a central parameter server, thus being more flexible with respect to the underlying connectivity \cite{koloskova2020unified}. This feature renders decentralized learning particularly appealing for future wireless networks with device-to-device communication. Several decentralized learning schemes over wireless networks have been proposed and analyzed  \cite{xing2020decentralized, ozfatura2020decentralized, Xing21-Simeone,Shi21}, highlighting the key role of over-the-air computation (AirComp) \cite{Amiri20-M} for low-latency training at the edge. 
Prior works have mainly considered wireless networks of reliable workers communicating in a fixed topology throughout the entire training procedure. Nevertheless, these assumptions are hardly met in practical systems, in which communication links can be intermittent or blocked, and devices may become temporarily unavailable due to computation impairments or energy saving reasons.
Asynchronous distributed training has been shown to mitigate the effect of stragglers (slow workers) \cite{Dutta21, nadiradze2019decentralized, adikari2020decentralized}. However, harnessing the potential benefits of asynchronism in decentralized learning over unreliable wireless networks remains elusive. 

In this paper, we propose an asynchronous implementation of decentralized stochastic gradient descent (DSGD) as a means to address the inherent communication and computation impairments of heterogeneous wireless networks. In particular, we study decentralized learning over a wireless network with a random time-varying communication topology, comprising unreliable devices that can become stragglers at any point of the learning process. To account for communication impairments, we propose a consensus strategy based on time-varying mixing matrices determined by the instantaneous network state. At the same time, we design the learning rates at the edge devices in such a way so as to preserve the stationary point of the original network objective in spite of the devices' heterogeneous computational capabilities. Finally, we provide a non-asymptotic convergence guarantee for the proposed algorithm, demonstrating that decentralized learning is possible even when outdated information from slow devices is used to locally train the models. Experimental results confirm our analysis and show that reusing stale gradient information can speed up convergence of asynchronous DSGD.

\section{System Model}
We consider a network consisting of $m$ wireless edge devices, in which each node $i$ is endowed with a local loss function $f_i:\mathbb{R}^d\to \mathbb{R}$ and local parameter estimate $\theta_i\in \mathbb{R}^d$. The network objective consists in minimizing the aggregate network loss subject to a consensus constraint
\begin{align}
\underset{\theta_1,\dots,\theta_m}{\text{minimize}}\ &f(\theta_1,\dots,\theta_m) := \frac{1}{m} \sum_{i=1}^m f_i(\theta_i)\label{problem} \\
\text{s.t.\quad } &\theta_1=\theta_2=\dots=\theta_m. \nonumber
\end{align}
This corresponds to the distributed empirical risk minimization problem whenever $f_i$ is a loss term over a local dataset. 
In the following, $f(\theta)$ denotes the network objective $f(\theta_1,\dots,\theta_m)\big|_{\theta_1=\dots=\theta_m=\theta}$ and $\bar{\theta}=1/m\sum^m_{i=1}\theta_i$.
To solve (\ref{problem}), we consider a DSGD algorithm according to which devices alternate between a local optimization based on gradient information (computation phase) and a communication phase.

\subsection{Computation model}
To locally optimize the model estimate $\theta_i$, we assume that each device can query a stochastic oracle satisfying the following properties.
\begin{assumption}
	At each node $i$, the gradient oracle $g_i(\theta)$ satisfies the following properties for all $\theta \in \mathbb{R}^d$
	\begin{itemize}
		\item  $\mathbb{E}[g_i(\theta)]=\nabla_\theta f_i(\theta)$ (unbiasedness)
		\item  $\mathbb{E}\norm{ g_i(\theta)-\nabla_\theta f_i(\theta)}^2\leq \sigma^2$ (bounded variance)
		\item  $\mathbb{E}\norm{g_i(\theta)}\leq G^2$ (bounded magnitude).
	\end{itemize}
	\label{ass:gradient}
\end{assumption}
We admit the existence of straggling nodes and that a random subset of devices can become inactive or postpone local optimization procedures, e.g., due to computation impairments or energy constraints. As a result, devices may join the communication phase and disseminate a model that has been updated using gradient information computed using previous model estimates, or a model that has not been updated at all from the previous iteration(s). Formally, at every optimization round $t$, the local update rule is
\begin{equation}
    \theta_i^{(t+\frac{1}{2})}=\begin{cases}
\theta_i^{(t)} , \quad \  \text{ if device $i$ is straggler at round $t$}\\
\theta_i^{(t)} -\eta^t_ig_i(\theta^{(t-\tau_i)}), \quad \text{otherwise}
\end{cases}
\label{eq:update}
\end{equation}
where $\eta^t_i$ is a local learning rate and the delay $\tau_i\geq 0$ accounts for the staleness of the gradient information at device $i$.

\subsection{Communication model}
The channel between any pair of device $i$ and $j$ follows a Rayleigh fading model. At every communication iteration $t$, devices can exchange information according to a connectivity graph $\mathcal{G}^{(t)}=(\mathcal{V},\mathcal{E}^{(t)})$, where $\mathcal{V} = \{1, 2, \dots,m\}$ indices the network nodes and $(i,j)\in\mathcal{E}^{(t)}$ if devices $i$ and $j$ can communicate during round $t$. We consider symmetric communication links; therefore the communication graph is undirected. While the connectivity graph is assumed to remain fixed within the optimization iteration, it may vary across optimization iterations due to deep fading, blockage, and/or synchronization failures. 

\section{Asynchronous Decentralized SGD}
\begin{figure}[t]
    \centering
    \includegraphics[width=0.86\columnwidth]{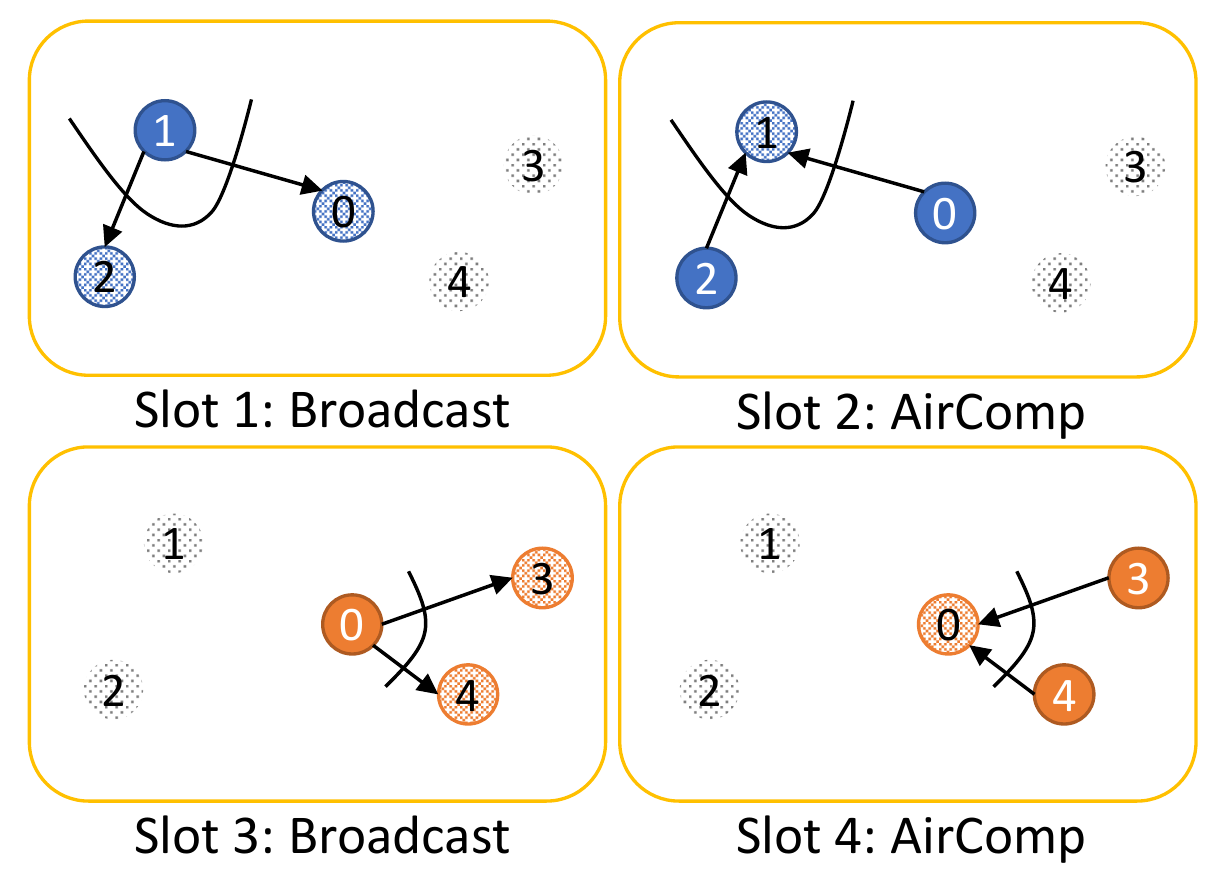}
    \caption{An example of the timeline for one training iteration composed of alternate Broadcast and AirComp slots.}
    \label{fig_scheduling}
\end{figure}

\begin{algorithm}
	\caption{Asynchronous Decentralized SGD}
	\begin{algorithmic}[1]
		\renewcommand{\algorithmicrequire}{\textbf{Input:}}
		\renewcommand{\algorithmicensure}{\textbf{Output:}}
		\REQUIRE $\theta_i^{(0)} = \mathbf{0} \in \mathbb{R}^d$
		\ENSURE  $\bar{\theta}^{(T)}$
		\FOR {$t$ in $[0, T]$}
    		\FORALL{non straggling devices}
    		\STATE update local model as (\ref{eq:update})
    		\ENDFOR
    		\STATE Determine matrix $W^{(t)}$ based on $\mathcal{G}^{(t)}$
    	\FOR {$s$ in $[1, S_t]$}
		\IF {$s \equiv 0 \pmod 2$}
		\STATE \textit{\# Broadcast phase}
		\FORALL{device $i$ scheduled in slot $s$}
	    \STATE Device $i$ transmits (\ref{eq:A_transmitted_broadcast})
		\STATE Each device $j\in\mathcal{N}^{(t)}_i$ receives (\ref{eq:A_received_broadcast})
		\STATE Each device $j\in\mathcal{N}^{(t)}_i$ estimates (\ref{eq:A_estimated_broadcast})
        \ENDFOR
		\ELSE
		\STATE \textit{\# AirComp Phase}
		\FORALL{star center  $i$ scheduled in slot $s$}
	    \STATE Each device $j\in\mathcal{N}^{(t)}_i$ transmits (\ref{eq:A_transmitted_broadcast})
		\STATE Device $i$ receives (\ref{eq:A_received_aircomp})
		\STATE Device $i$ estimates (\ref{eq:A_estimated_aircomp})
        \ENDFOR
		\ENDIF
		\ENDFOR
		\FORALL{device}
    		\STATE model consensus as in (\ref{eq:update_next_iter})
    		\ENDFOR
		\ENDFOR
	\end{algorithmic}
	\label{alg:async}
\end{algorithm}

The proposed asynchronous DSGD procedure, which takes into account both computation and communication failures, is detailed in Algorithm \ref{alg:async}. 

At the beginning of each training iteration $t$, non straggling devices update the local estimate $\theta^{(t)}_i$ according to (\ref{eq:update}) using a potentially outdated gradient information.
Subsequently, based on the current connectivity graph $\mathcal{G}^{(t)}=(\mathcal{V},\mathcal{E}^{(t)})$, devices agree on a symmetric and doubly stochastic mixing matrix $W^{(t)}$ using a Metropolis-Hastings weighting scheme \cite{xiao2006distributed}. The weights are very simple to compute and are amenable for distributed implementation. In particular, each device requires only knowledge of the degrees of its neighbors to determine the weights on its adjacent edges.

After that, it follows a communication phase in which devices exchange the updated estimates and employ a gossip scheme based on $W^{(t)}$. To leverage AirComp capabilities, devices employ analog transmission together with the scheduling scheme proposed in \cite{Xing21-Simeone}. Accordingly, the communication phase is divided into multiple pairs of communication slots. Each pair consists of an \emph{AirComp slot} and a \emph{broadcast slot} as illustrated in Fig.~\ref{fig_scheduling}.
During the AirComp slot $s$, the star center $i$ receives the superposition of the signals transmitted by its neighboring devices $\mathcal{N}^{(t)}(i)=\{j \in \mathcal{V}:(i,j) \in \mathcal{E}^{(t)}\}$. 
In particular, each scheduled node $j\in\mathcal{N}^{(t)}(i)$ transmits to the star center $i$
\begin{equation}
	x_{j}^{(s,t)} = \frac{\sqrt{\gamma^{(s,t)}_{i}}}{h^{(s,t)}_{i,j}}w^{(t)}_{i,j}\theta_{j}^{(t+\frac{1}{2})}
	\label{eq:A_transmitted_aircomp}
\end{equation}
where $h^{(s,t)}_{i,j}\in \mathbb{C}^d$ is the channel coefficient between user $i$ and $j$ during slot $s$, $\gamma^{(s,t)}_{i}\in \mathbb{R}$ is a power alignment coefficient, and $w^{(t)}_{i,j}$ is the $(i,j)$ entry of the mixing matrix ${W}$.
The star center $i$ receives the aggregated signal 
\begin{equation}
    y_i^{(s,t)} =\sum_{j\in \mathcal{N}(i)} h^{(s,t)}_{i,j}x^{(s,t)}_j + z_i^{(s,t)}
    \label{eq:A_received_aircomp}
\end{equation}
where $z_i^{(s,t)}\sim \mathcal{N}(0,\sigma_w\mathbb{1}_d)$ is a noise vector, and estimates the aggregated model as
\begin{equation}
    \hat{y}_i^{(s,t)} =\frac{ y_i^{(s,t)}}{\sqrt{\gamma^{(s,t)}_i}}=\sum_{j\in \mathcal{N}(i)} w^{(t)}_{i,j}\theta_{j}^{(t+\frac{1}{2})} + \frac{z_i^{(s,t)}}{\sqrt{\gamma^{(s,t)}_i}}.
    \label{eq:A_estimated_aircomp}
\end{equation}
On the other hand, during a broadcast slot $s$, scheduled node $i$ transmits using a power scaling factor $\alpha^{(s,t)}_{i}$ the signal
\begin{equation}
	x_{i}^{(s,t)}= \sqrt{\alpha^{(s,t)}_{i}}\theta_{i}^{(t+\frac{1}{2})}
	\label{eq:A_transmitted_broadcast}
\end{equation}
and all neighboring devices $j\in\mathcal{N}^{(t)}(i)$ receive
\begin{equation}
    y_j^{(s,t)} =h^{(s,t)}_{j,i}x^{(s,t)}_i + z_j^{(s,t)}
    \label{eq:A_received_broadcast}
\end{equation}
and estimate the updated model as
\begin{equation}
    \hat{y}_j^{(s,t)}=w^{(t)}_{j,i}\frac{ y_j^{(s,t)}}{\sqrt{\alpha^{(s,t)}_i}h^{(s,t)}_{j,i}}=w^{(t)}_{j,i}\left(\theta_{i}^{(t+\frac{1}{2})}+\frac{z^{(s,t)}_j}{\sqrt{\alpha_i}h_{j,i}}\right).
    \label{eq:A_estimated_broadcast}
\end{equation}
At the end of the communication phase, each node $i$ obtains the new estimate $\theta^{(t+1)}_i$ combining all received signals and using a consensus with step size $\zeta\in(0,1]$ 
\begin{equation}
    \theta_i^{(t+1)} = (1-\zeta)\theta_i^{(t+\frac{1}{2})} + \zeta \left\{\sum^m_{j=1}w^{(t)}_{i,j}\theta_{j}^{(t+\frac{1}{2})}+\tilde{n_i}^{(t)}\right\}
    \label{eq:update_next_iter}
\end{equation}
where $\tilde{n}_{i}^{(t)}\sim\mathcal{N}(0,\tilde{\sigma}^{(t)}_{w,i}\mathbb{1}_d)$ is a noise vector term that accounts for the aggregation of noise components during AirComp and broadcast transmissions at device $i$ during communication phase $t$.

\section{Convergence Analysis}
In this section, we study the effect of communication and computation failures on the asynchronous DGSD procedure and prove its convergence.

\subsection{Effect of Communication Failures}
Communication impairments amount for a random connectivity graph with an edge set that differs at each different optimization iteration. From an algorithmic perspective, random communication impairments result in DSGD with stochastic mixing matrices. A particular class of stochastic mixing matrices are those that satisfy the expected consensus property. 
\begin{definition}[Expected Consensus Rate \cite{koloskova2020unified}]
A random matrix ${W}\in\mathbb{R}^{m\times m}$ is said to satisfy the expected consensus with rate $p$ if for any $X\in \mathbb{R}^{d\times m}$
\[
\mathbb{E}_{W}\left[\norm{{W}X-\bar{X}}^2_F\right]\leq (1-p)\norm{X-\bar{X}}^2_F
\]
where $\bar{X}=X\frac{\mathbf{1}\mathbf{1}^T}{m}$ and the expectation is w.r.t. the random matrix ${W}$.
\end{definition}
\begin{lemma}\label{lem2}
If the event that the connectivity graph $\mathcal{G}^{(t)}$ is connected  at round $t$ has a probability $q>0$ and the Metropolis-Hastings weighting is used to generated the mixing $W^{(t)}$, the expected consensus rate is satisfied with rate $p=q\delta>0$, with $\delta$ being the expected consensus rate in case of a connected topology.
\end{lemma}
\begin{proof}
See Appendix \ref{subsect:proof_lem3}.
\end{proof}
If the expected consensus is satisfied, it is then possible to establish a convergent behavior for the estimates generated by the proposed algorithm.
\begin{lemma}[Consensus inequality]
	\label{lem:consensus}
	Under Assumption \ref{ass:gradient}, after $T$ iterations, decentralized SGD with a constant learning rate $\eta$ and consensus step size $\zeta$ satisfies 
	\begin{align*}
		\sum_{i=1}^{m}\norm{\theta^{(T)}-\bar{\theta}^{(T)}}_2\leq
		&\eta^2\frac{12mG^2}{(p\zeta)^2}+\zeta\frac{2}{p}\sum^m_{i=1}\sigma^2_{w,i}
	\end{align*}
where $\sigma^2_{w,i}=\max^T_{t=0}\mathbb{E}\norm{\tilde{n}^{(t)}_i}^2$.
\end{lemma}
\begin{proof}
See Appendix \ref{subsect:proof_lem4}.
\end{proof}

\begin{figure}
    \centering
    \includegraphics[width=0.91\linewidth]{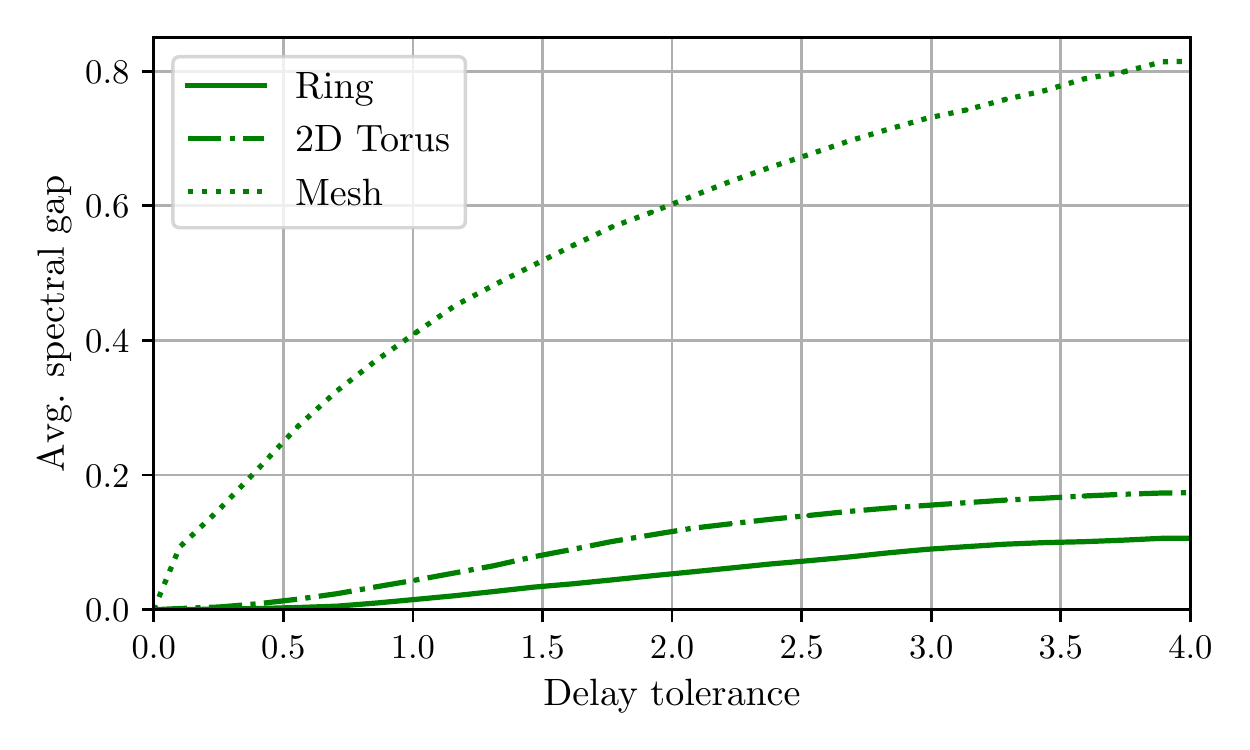}
    \caption{Average spectral gap under different delay constraints for mesh, ring, and two-dimensional torus topologies with 9 nodes. Each link is associated to a completion time $\sim Exp(1)$ and is dropped if it exceeds the delay tolerance value.}
    \label{fig:spectral_gap}
\end{figure}

Overall, communication failures amount to a reduced expected consensus rate compared to the scenario with perfect communication. At the same time, dropping users that are delayed and are unable to synchronize and perform AirComp, renders the communication protocol more flexible. For instance, in Fig.~\ref{fig:spectral_gap}, we consider a network of nine nodes organized according to different topologies and show the evolution of the average spectral gap of the mixing matrix with Metropolis-Hastings weights, whenever devices not satisfying a certain delay constraint are dropped. As expected, stricter delay requirements result in sparser effective communication graphs and mixing matrices with smaller spectral gaps.

\subsection{Effect of Computation Failures}
Random computation impairments make the group of devices that effectively update the model parameter vary over time. To account for this in the analysis, we introduce a virtual learning rate that is zero in case of failed computation. Namely, the learning rate at device $i$ during computation round $t$ becomes
\[
\tilde{\eta}_i^{(t)}=\begin{cases}
0 , \quad \  \text{ if $i$ is straggler at round $t$}\\
\eta^{(t)}_i, \quad \text{otherwise}
\end{cases}
\]
where $\eta^{(t)}_i$ is a specified learning rate value in case of successful computation. Furthermore, to ensure that the procedure converges to stationary points of the network objective even when edge devices have different computing capabilities, the expected learning rates have to be equalized. In particular, if $\mathbb{E}[\eta^{(t)}_i]=\eta, \ \forall i$, we have that stationary points are maintained in expectation, namely
\[
\sum_{i=1}^m\mathbb{E}[\tilde{\eta}_i^{(t)}]\nabla f_i(\theta)=0 \implies \nabla f(\theta)=0.
\]
Finally, the existence of straggling devices introduces asynchronicity in the decentralized optimization procedure. In particular, a device $i$ that fails at completing the gradient computation at a given optimization iteration is allowed to apply the result in a later one, without discarding the computation results. While we do not specify the delay distribution, we rather introduce the following assumption regarding the staleness of gradients.
\begin{assumption}
	For all iteration $t$, there exists a constant $\gamma\leq 1$ such that
	\begin{align*}
    	\mathbb{E}\norm{\nabla f(\bar{\theta}^{(t)})-\frac{\sum^m_{i=1}\nabla f_i(\theta^{(t-\tau_i)}_i)}{m}}^2&\leq\gamma 	\mathbb{E}\norm{\nabla f(\bar{\theta}^{(t)})}^2\\
    	&\hspace{-2em}+L^2\frac{\sum^m_{i=1}\mathbb{E}\norm{\theta_i^{(t)}-\bar{\theta}^{(t)}}^2}{m}.
	\end{align*}
	\label{ass:stale}
\end{assumption}
The above assumption is similar to the one in \cite{Dutta21} with an additional consensus error term. Note that the value of $\gamma$ is proportional to the staleness of the gradients and in case of perfect synchronization ($\gamma=0$) the bound amounts to a standard consensus error term.

\subsection{Convergence Guarantee}

In this subsection, we demonstrate the convergence of the decentralized optimization procedure to a stationary point of the problem (\ref{problem}).

\begin{theorem}\label{thm:stationary_cond}
Consider a network of unreliable communicating devices in which the expected consensus rate is satisfied with constant $p$ and each device can be a straggler with probability $\rho_i < 1$. If Assumptions \ref{ass:gradient} and \ref{ass:stale} are satisfied, asynchronous DSGD with constant learning rate $\eta_i=\min_j (1-\rho_j)/(\sqrt{4LT}(1-\rho_i))$ and consensus rate $\zeta=1/T^{3/8}$ satisfies the following stationary condition
\begin{align*}
\frac{1}{T}\sum^T_{t=1}\norm{\nabla f(\bar{\theta}^{(t)})}^2\leq& \frac{8\sqrt{L}(f(\bar{\theta}^{(T)})-f^*)}{\gamma' \rho_{min}\sqrt{T}}+\frac{3G^2L}{T^{1/4}p^2\gamma'}\\
&+\sqrt{\frac{L}{4T}}\frac{\sigma^2}{m\gamma'\min_j (1-\rho_j)}\\
&+\sum_{i=1}^m\frac{\sigma^2_{w,i}}{m\gamma'}\left(\frac{2L^2\gamma}{pT^{3/8}}+\frac{4L\sqrt{L}}{mT^{1/4} \rho_{min}}\right)\end{align*}
where $\gamma'=1-\gamma$, $\rho_{min}=\min_j (1-\rho_j)$ and $f^*=\min_{\theta\in \mathbb{R}^d}f(\theta)$.
\end{theorem}
\begin{proof}
See Appendix \ref{subsect:proof_stationary_cond}.
\end{proof}
The above theorem establishes a vanishing bound on the stationarity of the returned solution, which involves quantities related to both communication and computation impairments. In particular, the constant of the slowest vanishing terms $T^{-1/4}$ contains the term $p$ related to random connectivity, as well as $\gamma'$ and $\rho_{min}$ due to stragglers.

\section{Numerical Results}
\begin{figure}
	\centering
	\includegraphics[width=0.95\columnwidth]{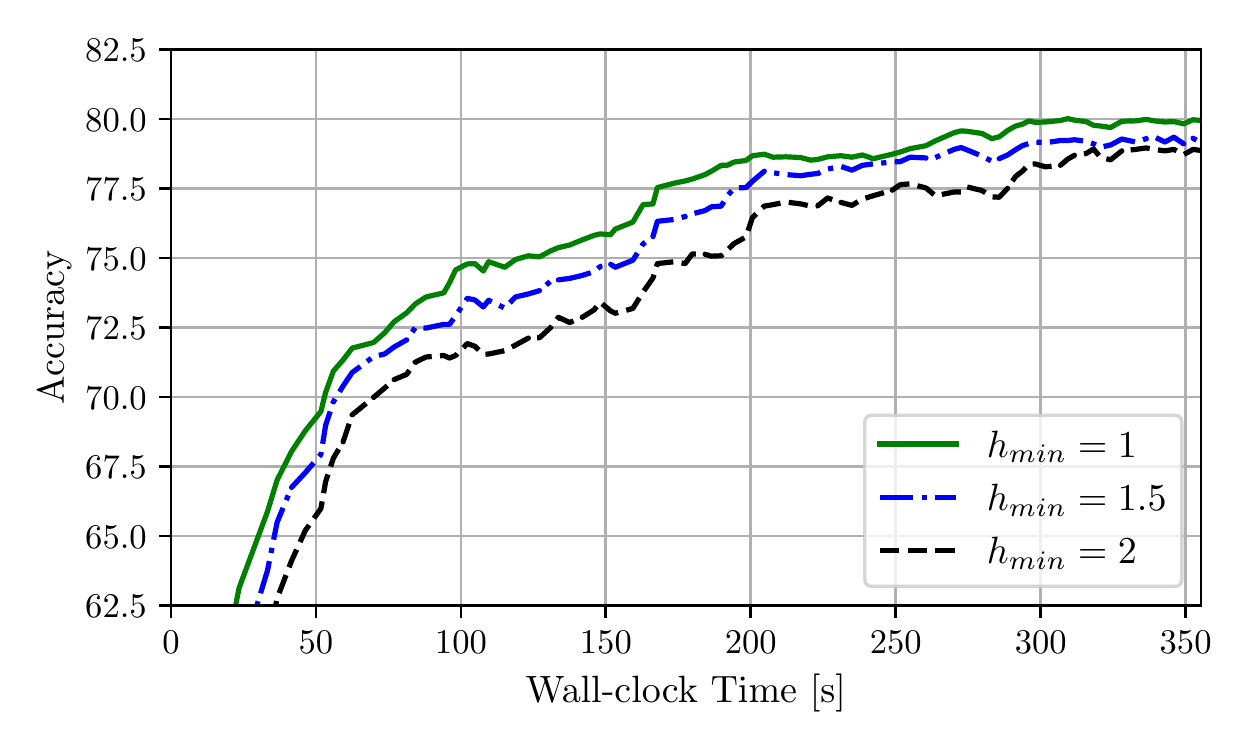} 
	\caption{Test accuracy versus time under different channel gain thresholds. Smaller thresholds result in larger average consensus rates and therefore in faster convergence.}\label{fig:comm_fail}
\end{figure}
The effectiveness of the proposed asynchronous DSGD scheme is assessed using a network of $m=15$ devices that collaboratively optimize the parameters of a convolutional neural network (CNN) for image classification with Fashion-MNIST. Gradients are calculated using batches of $16$ data samples and the performance is evaluated using a test set of 500 images. We model the channel gain between each device pair as Rayleigh fading and we assume a shifted exponential computation time at each device, i.e., $T_{comp}=T_{min}+Exp(\mu)$ with $T_{min}=0.25$s and $\mu=1$.
In Fig. \ref{fig:comm_fail}, nodes communicate only when the channel is in favorable conditions, i.e., when the channel gain exceeds a certain minimum threshold $h_{min}$. This allows to save energy; however, while higher threshold values result into lower average energy consumption, they also produce mixing matrices with smaller consensus rate, thus increasing the convergence time.

\begin{figure}
	\centering
	\includegraphics[width=0.94\columnwidth]{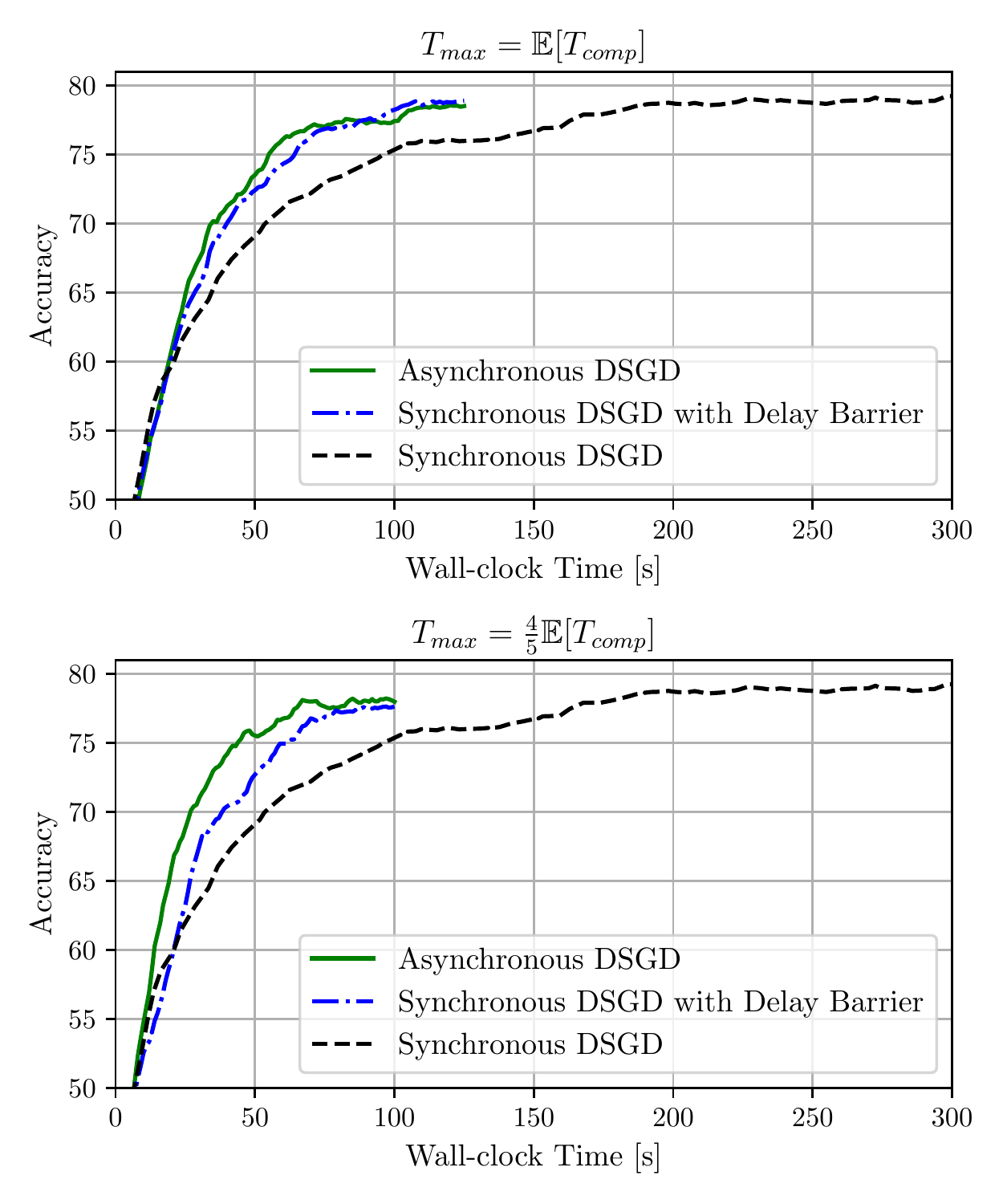} 
	\caption{Test accuracy for the asynchronous, synchronous with delay barrier, and synchronous schemes under two different values of $T_{max}$.}\label{fig:comp_fail}
\end{figure}

To study the effect of computation impairments, our proposed asynchronous learning algorithm is compared with: (i) \textit{synchronous DSGD}, which waits for all devices to finish their computations; and (ii) \textit{synchronous DSGD with a delay barrier $T_{max}$}, which discards computation from users that violate the maximum computing time. Compared to the latter, our asynchronous procedure allows for slow devices to reuse stale gradient computations during later iterations.
In Fig.\ref{fig:comp_fail}, we plot the evolution of the test accuracy of the aforementioned algorithms under two different values of $T_{max}$. For a moderate delay constraint $T_{max} = \mathbb{E}[T_{comp}]$, asynchronous DSGD and synchronous DSGD with delay barrier perform similarly as the fraction of slow users is modest. Nonetheless, imposing a delay constraint and discarding slow devices greatly reduces the training time compared to the synchronous DSGD case.
On the other hand, for a stringent delay requirement, $T_{max} = \frac{4}{5}\mathbb{E}[T_{comp}]$, reusing stale gradients turns out to be beneficial and the proposed asynchronous DSGD attains higher accuracy faster compared to the synchronous DSGD with a delay barrier.

\section{Conclusion}
In this work, we have proposed and analyzed an asynchronous implementation of DSGD, which enables decentralized optimization over realistic wireless networks with unreliable communication and heterogeneous devices in terms of computation capabilities.
We have studied the effect of both communication and computation failures on the training performance and proved non-asymptotic convergence guarantees for the proposed algorithm. The main takeaway is that reusing outdated gradient information from slow devices is beneficial in asynchronous decentralized learning.



\appendix

\subsection{Proof of Lemma~\ref{lem2}}
\label{subsect:proof_lem3}
Define the event $E^{(t)}:=\{\mathcal{G}^{(t)}\text{ is connected}\}$ and its complementary event $\bar{E}^{(t)}$. Whenever the Metropolis-Hasting weights are obtained from a connected graph, the resulting mixing matrix $W^{(t)}$ has a consensus rate greater than zero. Therefore, there exists $\delta>0$ such that
\begin{align*}
   \mathbb{E}_{W^{(t)}|E^{(t)}}\norm{W^{(t)} X-\bar{X}}^2_F\leq (1-\delta)\norm{W^{(t)} X-\bar{X}}^2_F
\end{align*}
It follows that, for any $X\in \mathbb{R}^{d\times m}$
\begin{align*}
    \mathbb{E}_{W^{(t)}}\norm{W^{(t)} X-\bar{X}}^2_F=&q\mathbb{E}_{W^{(t)}|E^{(t)}}\norm{W^{(t)} X-\bar{X}}^2_F\\
    &\hspace{-1em}+(1-q)\mathbb{E}_{W^{(t)}|\bar{E}^{(t)}}\norm{X-\bar{X}}^2_F\\
    \leq&q(1-\delta)\norm{W^{(t)} X-\bar{X}}^2_F\\
    &+(1-q)\norm{X-\bar{X}}^2_F\
\end{align*}
where we have lower bounded the consensus rate by zero in case of disconnected topologies. Grouping terms and having assumed $q>0$, we obtain that the expected consensus is satisfied with rate $(1-q\delta)>0$.

\subsection{Proof of Lemma~\ref{lem:consensus}}
 \label{subsect:proof_lem4}
Similarly to \cite{koloskova2019decentralized,Xing21-Simeone} we establish the following recursive inequality
\begin{align*}
		\sum_{i=1}^{m}\mathbb{E}\norm{\theta^{(t)}-\bar{\theta}^{(t)}}^2\leq&\left(1-\frac{p\zeta}{2}\right)\sum_{i=1}^{m}\mathbb{E}\norm{\theta^{(t-1)}-\bar{\theta}^{(t-1)}}^2\\
		&+\frac{\eta^2}{p\zeta}\left(6mG^2\right)+\zeta^2\sum^m_{i=1}\mathbb{E}\norm{\tilde{n}^{(t)}_i}^2.
\end{align*}
Defining $\sigma^2_{w,i}=\max^T_{t=0}\mathbb{E}\norm{\tilde{n}^{(t)}_i}^2$ and then solving the recursion we obtain the final expression.

\subsection{Proof of Theorem ~\ref{thm:stationary_cond}}
\label{subsect:proof_stationary_cond}
We denote stale gradients by $g_i(\tilde{\theta}^{(t)}_i)=g_i(\theta^{(t-\tau_i)}_i)$.
According to the update rule, at each iteration $t+1$, we have  
\[
\mathbb{E}[f(\bar{\theta}^{t+1})]=\mathbb{E}\left[f\left(\bar{\theta}^{t}-\frac{1}{m}\sum^m_{i=1}\left(\tilde{\eta}_i^{(t)} g_i(\tilde{\theta}^{(t)}_i)+\zeta \tilde{n}_i^{(t)}\right)\right)\right]
\]
where the expectation is w.r.t. the stochastic gradients, the communication noise $\Xi^{(t)}$, and the computation and communication failures at iteration $t+1$.
For an $L$-smooth objective function, we have
\begin{align*}
\mathbb{E}[f(\bar{\theta}^{(t+1)})]\leq f(\bar{\theta}^{(t)})\underbrace{-\frac{1}{m}\sum^m_{i=1}\left\langle\nabla f(\bar{\theta}^{(t)}),\mathbb{E}[\tilde{\eta}_i^{(t)} g_i(\tilde{\theta}^{(t)}_i))]\right\rangle}_{:=T_1}\\
+\underbrace{\frac{L}{2m^2}\mathbb{E}\norm{\sum^m_{i=1}\tilde{\eta}^{(t)}_ig_i(\tilde{\theta}^{(t)}_i))}^2}_{:=T_2} +\frac{L}{2m^2}\zeta^2\sum^m_{i=1}\mathbb{E}\norm{\tilde{n}_i^{(t)}}^2
\end{align*}
where we used the fact that the communication noise has zero mean and is independent across users.

Adding and subtracting $\nabla f_i(\bar{\theta}^{(t)})$ to each summand of $T_1$ and since $\mathbb{E}[\tilde{\eta}_i^{(t)} g_i(\tilde{\theta}^{(t)}_i)]=\eta\nabla f_i(\tilde{\theta}^{(t)}_i)$, with $\eta=\min_j (1-\rho_j)/(\sqrt{4LT})$,  we obtain
\begin{align*}
    T_1=&-\eta\left\langle\nabla f(\bar{\theta}^{(t)}),\frac{1}{m}\sum^m_{i=1}\nabla f_i(\tilde{\theta}^{(t)}_i)\right\rangle\\
    =& \frac{\eta}{2}\norm{\nabla f(\bar{\theta}^{(t)})-\frac{1}{m}\sum^m_{i=1}\nabla f_i(\tilde{\theta}^{(t)}_i)}^2\\
    &-\frac{\eta}{2}\norm{\nabla f(\bar{\theta}^{(t)})}^2-\frac{\eta}{2m^2}\norm{\sum^m_{i=1}\nabla f_i(\tilde{\theta}^{(t)}_i)}^2\\
    \leq&\frac{\eta\gamma}{2}\norm{\nabla f(\bar{\theta}^{(t)})}^2+\frac{\eta L^2}{2m}\sum^m_{i=1}\norm{\theta_i^{(t)}-\bar{\theta}^{(t)}}^2\\
    &-\frac{\eta}{2}\norm{\nabla f(\bar{\theta}^{(t)})}^2-\frac{\eta}{2m^2}\norm{\sum^m_{i=1}\nabla f_i(\tilde{\theta}^{(t)}_i)}^2\\
\end{align*}
where we have used the staleness assumption.
The last term can be bounded using the property of the stochastic gradient and the fact that $\tilde{\eta}^{(t)}_i\leq 1/(\sqrt{4LT})\leq 1/(\sqrt{4L})$ as
\begin{align*}
    T_2\leq&\frac{L}{2m^2}\mathbb{E}\norm{\sum^m_{i=1}\tilde{\eta}^{(t)}_i[g_i(\tilde{\theta}^{(t)}_i)-\nabla f_i(\tilde{\theta}^{(t)}_i)]}^2\\&+\frac{L}{2m^2}\mathbb{E}\norm{\sum^m_{i=1}\tilde{\eta}^{(t)}_i\nabla f_i(\tilde{\theta}^{(t)}_i)}^2\\
    \leq&\frac{\sigma^2}{8mT}+\frac{\eta}{8m^2}\mathbb{E}\norm{\sum^m_{i=1}\nabla f_i(\tilde{\theta}^{(t)}_i)}^2.
\end{align*}
Summing $T_1$ and $T_2$ we obtain
\begin{align*}
    T_1+T_2\leq&-\frac{\eta}{2}\left(1-\gamma\right)\norm{\nabla f(\bar{\theta}^{(t)})}^2+\frac{\sigma^2}{8mT}\\
    &+\frac{\eta L^2}{2m}\sum^m_{i=1}\norm{\theta_i^{(t)}-\bar{\theta}^{(t)}}^2\\
    & -\frac{\eta}{4m^2}\norm{\sum^m_{i=1}\nabla f_i(\tilde{\theta}^{(t)}_i)}^2.
\end{align*}
Defining $\gamma'=(1-\gamma)$, telescoping and taking expectations we obtain
\begin{align*}
\frac{1}{T}\sum^T_{t=1}\norm{\nabla f(\bar{\theta}^{(t)})}^2\leq& 2\frac{f(\bar{\theta}^0)-f(\bar{\theta}^T)}{\eta T\gamma'}+\frac{\sigma^2}{4\eta\gamma'mT}\\
&+\frac{1}{T}\sum^T_{t=1}\frac{ L^2}{m\gamma'}\sum^m_{i=1}\mathbb{E}\norm{\theta_i^{(t)}-\bar{\theta}^{(t)}}^2\\
&+\frac{1}{T}\sum^T_{t=1}\frac{L\zeta^2}{\eta  m^2\gamma'}\sum^m_{i=1}\mathbb{E}\norm{\tilde{n}_i^{(t)}}^2.
\end{align*}
Defining $\sigma^2_{w,i}=\max^T_{t=0}\mathbb{E}\norm{\tilde{n}^{(t)}_i}^2$ and bounding the consensus term by Lemma \ref{lem:consensus}, we obtain
\begin{align*}
\frac{1}{T}\sum^T_{t=1}\norm{\nabla f(\bar{\theta}^{(t)})}^2\leq& 2\frac{f(\bar{\theta}^0)-f(\bar{\theta}^T)}{\eta T\gamma'}\\
&+\frac{ L^2}{m\gamma'}\left(\eta^2\frac{12mG^2}{(p\zeta)^2}+\zeta\frac{2}{p}\sum^m_{i=1}\sigma^2_{w,i}\right)\\
&+\frac{\sigma^2}{4\eta\gamma'mT}+\frac{L\zeta^2}{\eta  m^2\gamma'}\sum_{i=1}^m\sigma^2_{w,i}.
\end{align*}
The final result is obtained setting $\eta=\frac{1}{\sqrt{4LT}}$ and $\zeta=\frac{1}{T^{3/8}}$.


\bibliographystyle{ieeetr}  
\bibliography{ref}

\end{document}